\documentclass[12pt]{article}

\usepackage{cite,epsfig,bbold}

\makeatletter

\def\section{\@startsection {section}{1}{\z@}{+3.0ex plus +1ex minus
  +.2ex}{2.3ex plus .2ex}{\normalsize\bf\boldmath}}

\def\mathswitch#1{\relax\ifmmode#1\else$#1$\fi}
\def\mathswitchr#1{\relax\ifmmode{\mathrm{#1}}\else$\mathrm{#1}$\fi}

\newcommand{\PZ}{\mathswitchr Z}

\newcommand{\Pe}{\mathswitchr e}

\newcommand{\Pt}{\mathswitchr t}

\newcommand{\me}{\mathswitch {m_\Pe}}

\newcommand{\mt}{\mathswitch {m_\Pt}}

\newcommand{\MZp}{\mathswitch {M_{\PZ'}}}
\newcommand{\GZp}{\mathswitch {\Gamma_{\PZ'}}}

\newcommand{\Br}{\mathswitchr{Br}}
\newcommand{\eqref}[1]{(\ref{#1})}
\newcommand{\lesim}{\,\raisebox{-.1ex}{$_{\textstyle <}\atop^{\textstyle\sim}$}\,}
\newcommand{\gesim}{\,\raisebox{-.3ex}{$_{\textstyle >}\atop^{\textstyle\sim}$}\,}
\newcommand{\Eslash}{{\not{\!\!E}}}

\begin{document}
\thispagestyle{empty}

\def\thefootnote{\fnsymbol{footnote}}

\begin{flushright}
FERMILAB--Pub--04/039--T\\
\end{flushright}

\vspace{1cm}

\begin{center}

{\Large\sc {\bf Weakly coupled neutral gauge bosons\\[1ex] at future linear colliders}}
\\[3.5em]
{\large
{\sc
A.~Freitas%
}
}

\vspace*{1cm}

{\sl
Fermi National Accelerator Laboratory, Batavia, IL 60510-500, USA
}

\end{center}

\vspace*{2.5cm}

\begin{abstract}
A weakly coupled new neutral gauge boson forms a narrow resonance that is hard
to discover directly in $e^+e^-$ collisions. However, if the gauge boson mass
is below the center-of-mass energy, it can be produced through processes where
the effective energy is reduced due to initial-state radiation and
beamstrahlung. It is shown that at a high-luminosity linear collider, such a
gauge boson can be searched for with very high sensitivity, leading to a
substantial improvement compared to existing limits from the Tevatron and also
extending beyond the expected reach of the LHC in most models. If a new vector
boson is discovered either at the Tevatron Run II, the LHC or the linear
collider, its properties can be determined at the linear collider with high
precision, thus helping to reveal origin of the new boson.
\end{abstract}

\def\thefootnote{\fnsymbol{footnote}}
\setcounter{page}{0}
\setcounter{footnote}{0}

\newpage


\section{Introduction}

Although the Standard Model, based on the gauge symmetry
SU(3)$\times$SU(2)$\times$U(1), is impressively successful in describing the
existing high-energy experimental results, it is expected to be only the
low-energy manifestation of a more fundamental theory, involving extended or
additional gauge groups. The additional gauge interactions may arise for
example in the framework of grand unified theories \cite{gut} or theories of
dynamical symmetry breaking \cite{tc}. 
Even if the fundamental gauge symmetry is broken at a scale far
beyond the electroweak breaking scale,
the breaking to the Standard
Model gauge group may occur in several steps and some subgroups may remain
unbroken at a scale not far from the electroweak scale. Therefore it is
interesting to study a possible additional neutral gauge boson $Z'$ with mass
below 1 TeV. Such a new gauge boson with a mass as low as the order of the $Z$
mass is still in accordance with all experimental bounds if its couplings to
the Standard Model fermions are very weak \cite{leike:99}.

Current limits from searches at the Tevatron \cite{TeVcur} exclude a
"sequential $Z'$ boson" (i.e. a $Z'$ boson that has the same couplings as the
Standard Model $Z$ boson) with mass below 780 GeV. For smaller masses, the
sensitivity towards smaller couplings increases. For $Z'$ masses as low as 300
GeV or less, the Tevatron can exclude $Z'$ bosons with production
cross-sections that are about 100 times smaller than for a sequential $Z'$
boson.

The LEP experiments
could also place limits on neutral gauge bosons, but the LEP data has not been
analyzed for weakly coupled $Z'$ bosons. It is expected that the limits
obtainable from LEP, normalized to a sequential $Z'$ boson, will be slightly
better but comparable to the Tevatron limits for $Z'$ bosons with masses below
$210$ GeV \cite{bogdan:03}.

The searches for a relatively light weakly coupled $Z'$ boson could be improved
by a future $e^+e^-$ high-energy linear collider with high luminosity
\cite{lc}, if the $Z'$ boson couples to electrons. This report estimates the
reach of a linear collider for $Z'$ masses of the order of 1 TeV, focusing on
the case that the mass of the $Z'$ boson is below the $e^+e^-$ center-of-mass
energy. The findings are compared to the reach of the LHC for neutral gauge
bosons. In addition to the discovery potential of an $e^+e^-$ collider, the
capabilities of determining the couplings of the $Z'$ boson are investigated.
For the most part of the analysis, no assumptions are made about the specific
model structure that gives rise to the $Z'$ boson. However, as an example, one
interesting class of models studied in Ref.~\cite{bogdan:03}, where the
Standard Model is extended by only one additional U(1) gauge group and no extra
fermions, is analyzed in more detail.

\section{Direct $Z'$ production} \label{direct}

In the general setup for this study it is assumed that mixing effects between
the $Z$ and $Z'$ bosons, introduced through off-diagonal entries in the neutral
gauge-boson mass matrix, are negligible. When both the $Z$ and $Z'$ bosons
couple to the Standard Model fermions, there will also be kinetic mixing
generated by loop contributions at higher orders. These mixing effects,
however, can always be rotated away for on-shell momenta of the gauge bosons by
diagonalizing the higher-order propagator matrix. For small $Z$--$Z'$ mixing,
no important bounds on the $Z'$ boson arise from indirect constraints from
$Z$-pole data.

In this case, the most stringent bounds are obtained from direct $Z'$
production. If the coupling $g_{Z'ff}$ of the $Z'$ boson to fermions (or other
possible decay products) is small, it will form a narrow resonance that is hard
to discover in the process $e^+e^- \to f \bar{f}$. Very stringent constraints
can be obtained for values of the $Z'$ mass 
close to the center-of-mass
energy
, $\MZp \approx \sqrt{s}$. When the width is smaller than the
detector resolution and the beam energy spread, 
the production cross-section for $\MZp = \sqrt{s}$ can be expressed as
\begin{equation}
\int {\rm d}(\sqrt{s}) \;
\sigma[e^+e^- \to Z' \to f\bar{f}] = \frac{6\pi^2\GZp}{\MZp^2}
\, \Br(Z' \to e^+e^-) \, \Br(Z' \to f \bar{f}), \label{eq:onres}
\end{equation}
where the integration is performed over one energy bin.
For a narrow $Z'$ resonance away from the nominal center-of-mass energy, $|\MZp
- \sqrt{s}| \gg \GZp$, the sensitivity of the process $e^+e^- \to f \bar{f}$
quickly decreases. As pointed out in \cite{leike:99,bogdan:03}, the most
stringent constraints for $\MZp < \sqrt{s}$ are obtained from the case where the
invariant mass of the $f \bar{f}$--system
is reduced by initial-state radiation, so that the $Z'$ boson can still be
produced on-shell,
\begin{equation}
e^+e^- \to Z' + n\gamma \to f \bar{f} + n\gamma. \label{eq:radreturn}
\end{equation}
The leading initial-state radiation effects due to large logarithms, $L = \log
s/\me^2$, can be included using
the structure-function approach \cite{structf},
\begin{equation}
\sigma[e^+e^- \to f \bar{f} + n\gamma](s) = \int_0^1 {\rm d}x_+ \int_0^1 {\rm d}x_-
  \, G_{ee}(x_+,s) \, G_{ee}(x_-,s) \;
  \sigma[e^+e^- \to f \bar{f}](s x_+ x_-),
\end{equation}
with the structure functions up to order ${\cal O}(\alpha^2 L^2)$%
\footnote{The leading-logarithmic QED structure functions are known to higher
orders beyond ${\cal O}(\alpha^2 L^2)$ \cite{LLnew}, which is however not relevant for the level of precision of
this study.}
and including soft-photon exponentiation given by \cite{LLold}
\begin{equation}
\begin{array}{@{}r@{\;}c@{\;}l@{}}
G_{ee}(x,Q^2) &=& \displaystyle
  \frac{\zeta_\alpha \, (1-x)^{\zeta_\alpha-1}}{\Gamma(1+\zeta_\alpha)}
  \, e^{-\gamma_{\rm E} \zeta_\alpha + 3\alpha L/4 \pi}
 - \frac{\alpha}{2\pi} L \, (1+x) \\[2ex]
&-&  \displaystyle
  \frac{1}{2} \left( \frac{\alpha}{2\pi} \right)^2 L^2
        \left[ \frac{1+3x^2}{1-x} \log x + 4(1+x)\,\log(1-x) + 5 + x \right],
\end{array}
\end{equation}
where $\zeta_\alpha = \alpha (L-1)/\pi$ and $\gamma_{\rm E} \approx
0.577$ is the Euler constant. The additional photons usually escape in the
direction of the beam pipe and will not be considered for the signal signature.

Besides initial-state radiation, beamstrahlung plays an important role for
high-luminosity $e^+e^-$ colliders. It also leads to an effective reduction of
the invariant mass of the hard scattering process. In this work beamstrahlung
effects have been included using the program {\sl Circe} \cite{circe} for {\sc
Tesla} design parameters \cite{tesla}. This program provides effective
parameterizations of detailed multi-body simulations obtained from the code {\sl
Guinea-Pig} \cite{guinea-pig}.

Both photonic initial-state radiation and beamstrahlung contribute to the
effective cross-section for $Z'$ production. The effect from beamstrahlung
becomes particularly important for $Z'$ masses close to the center-of-mass
energy, contributing up to about 50\% of the cross-section.
For smaller values of the ratio $\MZp/\sqrt{s}$, the contribution from
beamstrahlung reduces to a few percent compared to initial-state radiation.
The relative importance of beamstrahlung 
also grows with increasing beam energy.
For this study, the beamstrahlung effects have been investigated only for the
{\sc Tesla} accelerator proposal. Generally speaking,
the NLC/JLC \cite{nlc} and CLIC \cite{clic} linear collider designs 
generate slightly larger beamstrahlung effects than the {\sc Tesla} design,
especially towards the lower end of the beam energy spectrum, i.e. for
small effective beam energies (see e.g. \cite{circe,napoly:92}). Therefore 
the search prospects for $Z'$ bosons with $\MZp \ll \sqrt{s}$ are slightly
better for the NLC/JLC and CLIC designs as opposed to the {\sc Tesla} design.
However, since in this region the cross-section is dominantly generated by
initial-state radiation, the general results of this study
remain valid also for these other accelerator designs.

The main background to the signal process \eqref{eq:radreturn} arises from
fermion pair production through s-channel photon or $Z$ boson exchange,
\begin{equation}
e^+e^- \to \gamma^*/Z^* + n\gamma \to f\bar{f} + n\gamma. \label{eq:bk}
\end{equation}
Additional backgrounds arise from processes with higher particle multiplicity
in the final state, for example leptons in conjunction with additional
neutrinos or four-quark final states from $W$-boson or tau pair production.
For a realistic experimental analysis, the four-fermion backgrounds may play a
non-negligible role and should be taken into account.
Nevertheless,
since these backgrounds are sub-dominant with respect to the process
\eqref{eq:bk}, they will not modify the discovery reach for a $Z'$ boson in a
significant way and are thus not included in the present study.

Both the signal \eqref{eq:radreturn} and background \eqref{eq:bk} contributions
have been calculated using Monte-Carlo techniques including the initial-state
radiation and beamstrahlung effects. 
No detector effects or experimental acceptances are taken into account in this
study.
The coverage of the $Z'$ parameter space could be improved by running
at different center-of-mass energies. Here the following scenarios are
considered: (i) near the $W$-boson threshold with $\sqrt{s} = 170$ GeV and
integrated luminosity ${\cal L} = 50 {\rm \ fb}^{-1}$ \cite{wilson:01}, (ii) near
the $t$-quark threshold with $\sqrt{s} = 350$ GeV and
${\cal L} = 100 {\rm \ fb}^{-1}$ \cite{tt}, (iii) a
base-line high-energy design with $\sqrt{s} = 500$ GeV and
${\cal L} = 500 {\rm \ fb}^{-1}$ \cite{lc}, and (iv) an upgraded high-energy version
with $\sqrt{s} = 1000$ GeV and
${\cal L} = 1000 {\rm \ fb}^{-1}$.

\begin{figure}[p]
\epsfig{figure=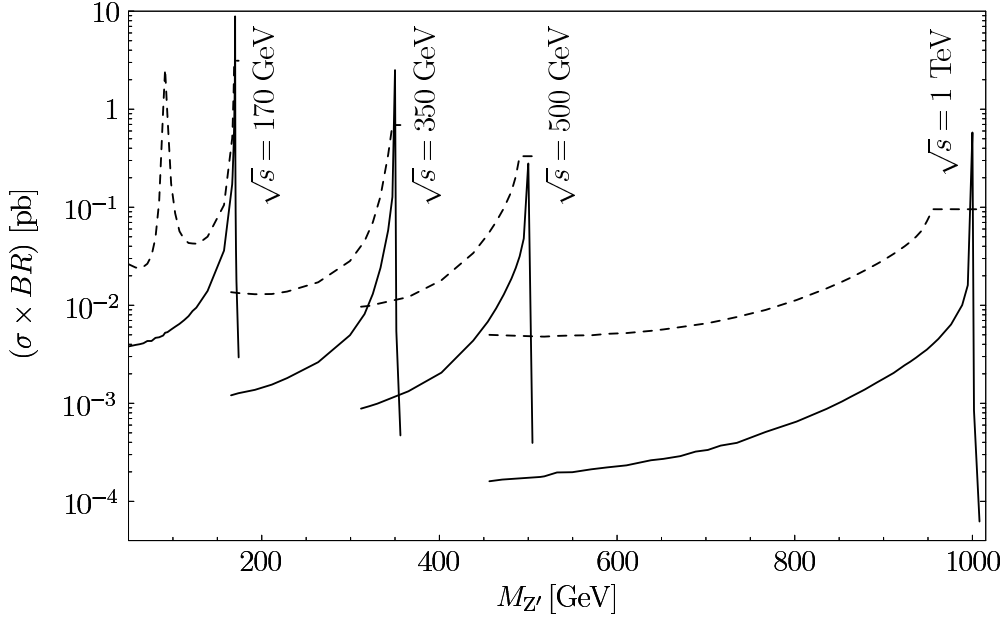, width=16cm, bb=0 412 509 660}
\vspace{-2ex}
\caption{Signal cross-section times branching ratio (solid lines) in the $\mu^+\mu^-$ channel 
for a sample $Z'$ model with
couplings that are 30 times smaller than for a sequential $Z'$ boson. The
cross-sections are shown as a function of the mass of the $Z'$ boson, $\MZp$,
for different collider energies. Also shown is the background cross-section
(dashed lines) in a bin in the $\mu$-pair invariant mass with a bin size given by
eq.~\eqref{eq:muenres}.}
\label{fg:Xsecm}
\end{figure}

As a starting point, the
cleanest decay channel, $Z' \to \mu^+ \mu^-$, will be considered.
The presence of a $Z'$ boson
would show up as a resonance peak in the $\mu$-pair invariant mass spectrum.
For the detection of a narrow resonance, a good momentum resolution is therefore
important. Here the parameters of the {\sc Tesla} study \cite{tesladet} are
taken, which envisages a momentum resolution of $\Delta(1/p) = 5 \times 10^{-5}
{\rm \ GeV}^{-1}$ in the central tracker region. For a muon pair, this number
needs to be multiplied by two. Due to the boost from the initial-state photon,
the muons are not always produced back-to-back. 
Therefore a conservative additional factor of two is included to accommodate
for possible topological effects. Considering that the maximum energy of the
muons is $\sqrt{s}/2$, one obtains for the resolution of the $\mu$-pair
invariant mass:
\begin{equation}
\Delta E_{\mu^+\mu^-} = 5 \times 10^{-5} {\rm \ GeV}^{-1} \times s.
\label{eq:muenres}
\end{equation}
To compute the significance of a possible $Z'$ signal, the invariant mass
spectrum of the $\mu$ pair is divided into bins of the size given by
eq.~\eqref{eq:muenres}. Due to the narrow width of weakly coupled $Z'$ boson,
the signal will appear as an excess in a single bin.
The expected signal cross-sections as a function of the $Z'$ mass
are exemplified in Fig.~\ref{fg:Xsecm} for a $Z'$ boson with couplings that are
30 times smaller compared to the Standard Model $Z$ boson,
$g_{Z'ff}/(g_{Zf\bar{f}})_{\raisebox{-1mm}{\scriptsize SM}} = 30$. In this case
the signal cross-section is reduced by a factor 900 relative to a sequential
$Z'$ boson. Also shown is the
background level from the processes eq.~\eqref{eq:bk} in the given bin.%
\begin{figure}[p]
\epsfig{figure=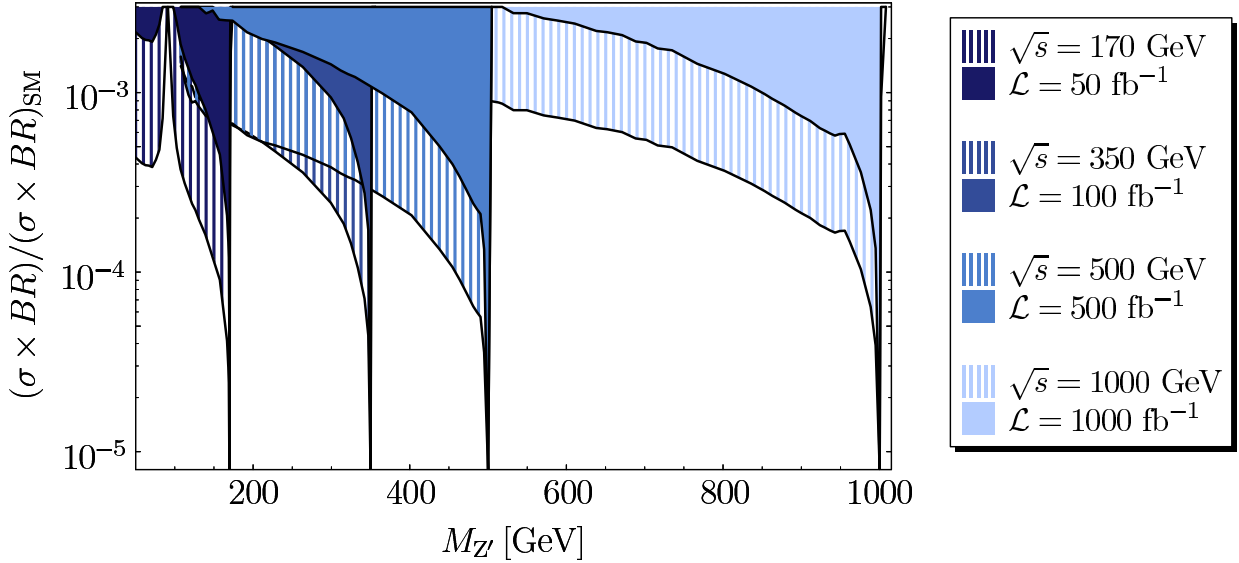, width=\textwidth, bb=75 460 515 660}
\vspace{-3ex}
\caption{Projected sensitivity of a future $e^+e^-$ collider for exclusion at
95\% confidence level (hatched regions) and $5\sigma$ discovery (solid
regions) of an additional neutral gauge boson in the $\mu^+\mu^-$ channel. 
Shown is the reach in terms of the product of production cross-section and
branching ratio, normalized to the value for a Standard Model $Z$ boson, 
as a function of the gauge boson mass, $\MZp$, for various collider energies.}
\label{fg:mumu}
\end{figure}
With luminosities of a few hundred fb$^{-1}$ one expects signal rates of
1000--10000 events for the depicted example, with background levels that are
about one order of magnitude larger.
The $Z'$ boson is searched for by counting the events 
in each invariant mass bin and looking for an excess in one bin while vetoing
significant deviations from the background expectation in the other bins.
In order to determine the statistical significance that the excess in one bin 
is not a statistical fluctuation, a $\chi^2$ test over all bins is performed.

For a viable event, the two muons are required to be in the main region of the
detector, $|\cos \theta_{\mu^\pm}| < 0.94$, where $\theta_{\mu^\pm}$ is the polar
angle of the $\mu^\pm$. Taking into account these constraints, the projected
reach of a linear collider for a $Z'$ boson is shown in Fig.~\ref{fg:mumu}.
The results are presented in
terms of the product of production cross-section times branching ratio
Br$(Z'\to\mu^+ \mu^-)$, normalized to the case where the $Z'$ couplings are
identical to the Standard Model $Z$ boson.%
\begin{figure}[p]
\epsfig{figure=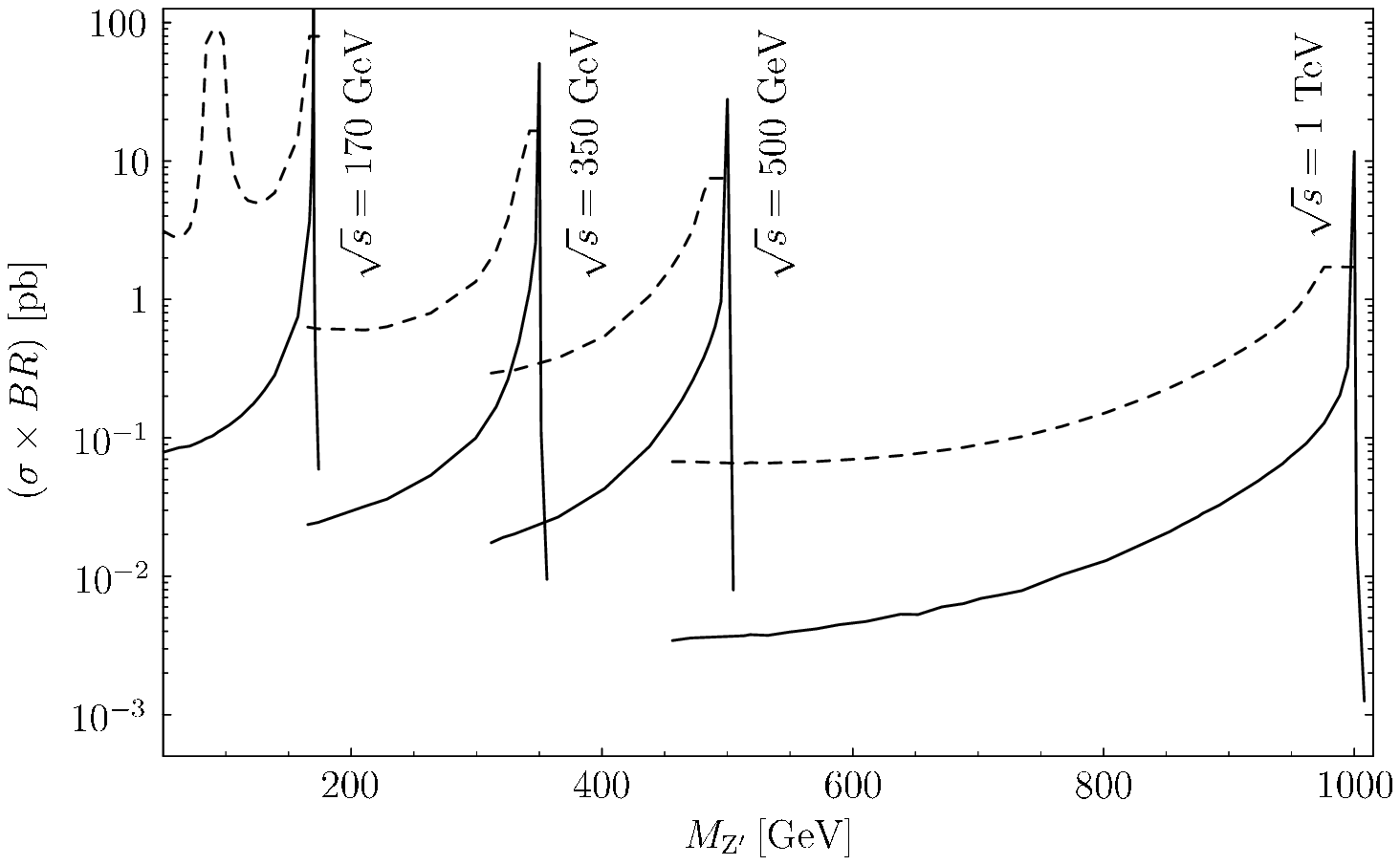, width=16cm, bb=0 416 509 660}
\vspace{-2ex}
\caption{Signal cross-section times branching ratio (solid lines) in the di-jet channel 
for a sample $Z'$ model with
couplings that are 30 times smaller than for a sequential $Z'$ boson. The
cross-sections are shown as a function of the mass of the $Z'$ boson, $\MZp$,
for different collider energies. Also shown is the background cross-section
(dashed lines) in a bin in the di-jet invariant mass with a bin size given by
eq.~\eqref{eq:qenres}.}
\label{fg:Xsecq}
\end{figure}
\begin{figure}[p]
\epsfig{figure=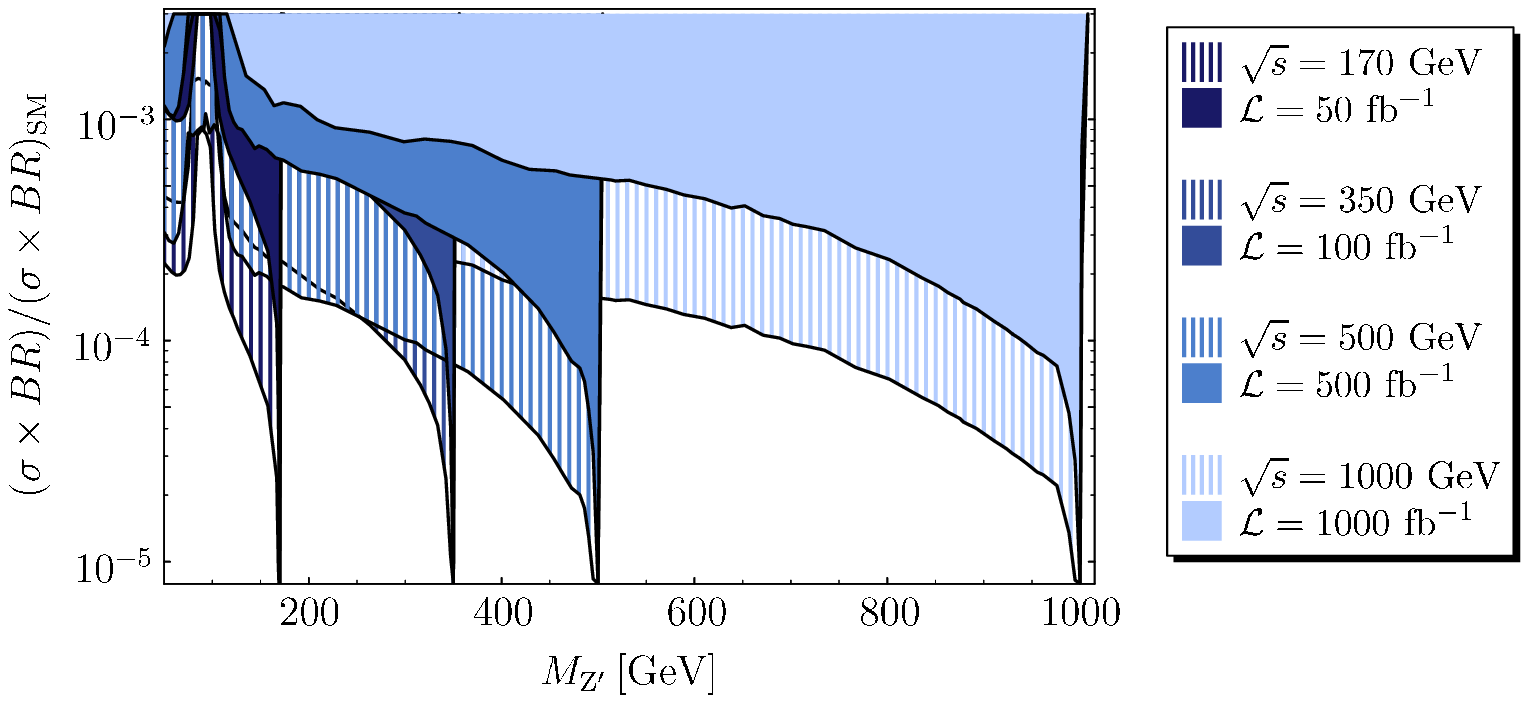, width=\textwidth, bb=75 460 515 660}
\vspace{-3ex}
\caption{Projected sensitivity of a future $e^+e^-$ collider for exclusion at
95\% confidence level (hatched regions) and $5\sigma$ discovery (solid
regions) of an additional neutral gauge boson in the jet-jet channel. 
As in Fig.~\ref{fg:mumu}.}
\label{fg:qq}
\end{figure}
As evident from the figure, by combining information from the various collider
energies, a $Z'$ boson with a signal rate that is about three orders of
magnitude smaller than for a gauge boson with Standard Model $Z$ couplings can
be found throughout the range 50 GeV $< \MZp <$ 1 TeV, except near the $Z$
resonance.

The sensitivity can be increased by including other final-state channels. For
the $Z' \to e^+e^-$ channel, a similar sensitivity as for the $\mu^+\mu^-$
channel is expected. While the $e^+e^-$ channel is plagued by larger background
contributions due to t-channel photon or $Z$-boson exchange, most of this
background is peaked in the forward or backward regions of the detector. By
restricting the signal to the central detector region, the background levels for
the $e^+e^-$ and $\mu^+\mu^-$ channel roughly comparable. 
Under the assumption of lepton universality, the two lepton channels can be
statistically combined, thus improving the
limits in Fig.~\ref{fg:mumu} by about a factor $\sqrt{2}$.
The inclusion of the decay channel into taus, $Z' \to \tau^+\tau^-$ is more
demanding, because of the missing energy carried away by the neutrinos in the
tau decays, and will not be considered here.

Depending on the branching ratios of the $Z'$,  the inclusion of the decay
channels into hadrons, $Z' \to q\bar{q},$ $q \neq t$, can lead to stronger
limits than the leptonic decay channels. Here only the decays into the five
light quark flavors are considered, which are characterized by a two-jet
signature. Due to the different decay signature of the top-quark, its
contribution would have to be treated separately. However, a statistical
combination with the light-quark channels is only possible using some
assumptions about the underlying $Z'$ model and will not be performed here.
The resolution for the
invariant mass spectrum of a jet pair is governed by the energy resolution of
the hadronic calorimeter. The jet energy resolution for a jet with energy $E$ 
is expected to be $\Delta E/E = 35\%/\sqrt{E/{\rm GeV}} \oplus 3\%
$ \cite{tesladet},
where $\oplus$ indicates quadratic combination of the errors. By using again the
maximal energy of the final-state jets, $E_{\rm max}=\sqrt{s}/2$, one obtains the following
(conservative) estimate for the resolution of the di-jet invariant mass spectrum,
\begin{equation}
\Delta E_{jj} = 2 \, \sqrt{0.06125 {\rm \ GeV} \times \sqrt{s} + 0.000225 \, s}.
\label{eq:qenres}
\end{equation}
Similar to the muon channel, the signal is characterized by a narrow peak
showing up as an excess in a single bin in a binned analysis of the 
di-jet invariant mass spectrum. As an example, Fig.~\ref{fg:Xsecq} illustrates
the expected signal and background cross-sections for a $Z'$ boson with
couplings that are 30 times smaller compared to the Standard Model $Z$ boson
and a bin size given by eq.~\eqref{eq:qenres}.

Fig.~\ref{fg:qq} shows the expected sensitivity of a linear collider for a $Z'$
boson in the hadronic decay channel. The improvement with respect to the
$\mu^+\mu^-$ channel is mainly a result of the normalization to the branching
ratios of the Standard Model $Z$ boson, which has relatively small couplings to
the charged leptons, $\Br(Z\to\mu^+\mu^-) \approx 3.4$\%, but large couplings
to the quarks, $\Br(Z\to q\bar{q}) \approx 69$\%.

It should be noted that besides the possibility to search for $Z'$ boson in
the decay channels into charged leptons and quarks, it is also possible to
search for ``invisible'' $Z'$ bosons, which primarily decay into neutrinos or
other weakly interacting particles. The best limits for
invisible $Z'$ searches can be obtained from the process $e^+e^- \to \gamma +
{\rm missing\ energy}$, where a hard photon is required for tagging
\cite{invZ}. The presence of a $Z'$ boson would show up as a resonance peak in
the energy distribution of the photon. In contrast to the radiative-return
method for visibly decaying $Z'$ bosons, as discussed above, here the photon is
not allowed to be in the kinematical region collinear to the beam pipe. As a
consequence of this, the projected limits will be weakened by roughly a factor
$L = \log s/\me^2$ compared to the discovery limits shown in
Figs.~\ref{fg:mumu}, \ref{fg:qq}. \label{inv}

One of the most intriguing cases for the searches outlined above is the
situation where only one $Z'$ gauge boson but no evidence
for other particles beyond the particle content of the Standard Model is found
at future colliders. This situation has been studied in detail in
Ref.~\cite{bogdan:03}. The full gauge group of this model SU(3)$_{\rm C}$
$\times$ SU(2)$_{\rm L}$ $\times$ U(1)$_{\rm Y}$ $\times$  U(1)$_{\rm Z'}$
extends the Standard Model gauge symmetry by an additional symmetry group
U(1)$_{\rm Z'}$, that is spontaneously broken by the vacuum expectation value
of a new scalar field\footnote{Note that the scalar $\phi$ can easily
escape observation when it is charged only under the additional gauge group
U(1)$_{\rm Z'}$, so that it couples to the Standard Model particles 
only through small mixing effects.}, $\phi$. Under the assumption that mixing 
effects between the
$Z$ and $Z'$ bosons are small, the $Z'$ boson is directly associated with the
U(1)$_{\rm Z'}$ group.
The quantum numbers of the Standard Model quarks,
$q_{\rm L}$, $u_{\rm R}$, $d_{\rm R}$, and leptons, $l_{\rm L}$, $e_{\rm R}$,
and the Higgs doublet $H$ are tightly constrained by anomaly cancellation
(assuming generation-independent charges), leaving at most two independent
parameters $z_u$ and $z_q$, see Tab.~\ref{tab:charges}. In general, these are
further constrained depending on the number of right-handed neutrinos charged
under U(1)$_{\rm Z'}$. However, for the following discussion the right-handed
neutrinos are assumed to be heavy and therefore irrelevant.

\begin{table}[tb]
\renewcommand{\arraystretch}{1.2}
\centering
\begin{tabular}{|c||c|c|c|c|}
\hline
 & SU(3)$_{\rm C}$ & SU(2)$_{\rm L}$ & U(1)$_{\rm Y}
$ & U(1)$_{\rm Z'}$ \\
\hline\hline
$\;q_{\rm L}\;$ & $\mathbb{3}$ & $\mathbb{2}$ & 1/3 & $z_q$ \\
$u_{\rm R}$ & $\mathbb{3}$ & $\mathbb{1}$ & 4/3 & $z_u$ \\
$d_{\rm R}$ & $\mathbb{3}$ & $\mathbb{1}$ & -2/3 & $2z_q-z_u$ \\
\hline
$l_{\rm L}$ & $\mathbb{1}$ & $\mathbb{2}$ & -1 & $-3z_q$ \\
$e_{\rm R}$ & $\mathbb{1}$ & $\mathbb{1}$ & -2 & $-2z_q-z_u$ \\
\hline
$H$ & $\mathbb{1}$ & $\mathbb{2}$ & 1 & $-z_q+z_u$\\
$\phi$ & $\mathbb{1}$ & $\mathbb{1}$ & 0 & 1 \\
\hline
\end{tabular}
\caption{The most general allowed charge assignments for fermions and scalars
under the extended gauge symmetry SU(3)$_{\rm C}$ $\times$ SU(2)$_{\rm L}$
$\times$ U(1)$_{\rm Y}$ $\times$ U(1)$_{\rm Z'}$ when requiring anomaly
cancellation and generation-blind couplings \cite{bogdan:03}.}
\label{tab:charges}
\end{table}

\begin{figure}[tb]
\centering
\epsfig{figure=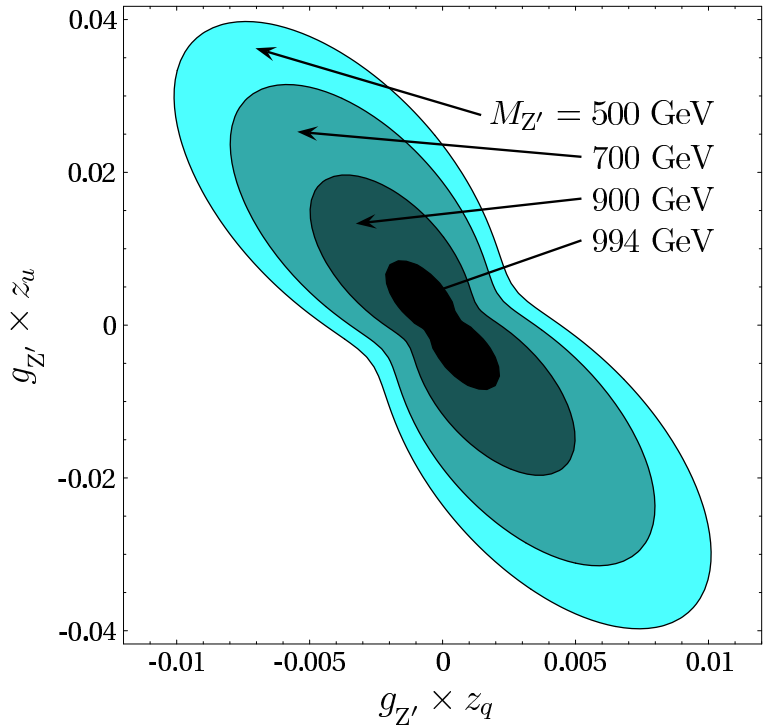, width=10cm, bb=19 445 274 682}
\caption{Discovery reach of a $e^+e^-$ collider in the $\mu^+\mu^-$
channel for $\sqrt{s} = 1$ TeV in terms of the charges $z_u$ and $z_q$,
multiplied by the gauge coupling $g_{\rm Z'}$ of the extra U(1) group.
The shaded areas indicate the remaining regions where a discovery is
\emph{not} possible for different values of the $Z'$
mass.}
\label{fg:zuzq}
\end{figure}

Neglecting the masses of the light fermions $f$, $f \neq t$,
the leading-order partial and total decay widths of the $Z'$ boson read
\begin{equation}
\begin{array}{@{}r@{\;}c@{\;}ll}
\Gamma[Z' \to \mu^+\mu^-] &=& \displaystyle
\frac{g_{\rm Z'}^2}{96\pi} \MZp (13 z_q^2 + 4 z_q z_u + z_u^2), \\[2ex]
\displaystyle \sum_{q\neq t} \Gamma[Z' \to q \bar{q}] &=& \displaystyle
\frac{g_{\rm Z'}^2}{32\pi} \MZp (17 z_q^2 - 12 z_q z_u + 5 z_u^2), \\[2ex]
\Gamma_{\rm Z',tot} &=& \displaystyle
\frac{g_{\rm Z'}^2}{32\pi} \MZp (39 z_q^2 - 8 z_q z_u + 6 z_u^2
  + \delta\Gamma_{\rm t}),
\label{eq:Gam}
\end{array}
\end{equation}
$$
\delta \Gamma_{\rm t} = \left\{ \begin{array}{cl}
\displaystyle
  \sqrt{1-4\frac{\mt^2}{\MZp^2}} \left[(z_q + z_u)^2 
    \left( \frac{1}{2} + \frac{\mt^2}{\MZp^2} \right) + (z_q - z_u)^2
    \left( \frac{1}{2} - 2 \frac{\mt^2}{\MZp^2} \right) \right]
    & \mbox{ for } \MZp > 2 \mt \\[1.5em]
    0 & \mbox{ for } \MZp < 2 \mt,
  \end{array} \right. 
$$
where $g_{\rm Z'}$ is the gauge coupling associated with the gauge group
U(1)$_{\rm Z'}$. As evident from \eqref{eq:Gam}, the production and decay of
the $Z'$ boson is fully described by three unknown quantities, the mass $\MZp$
and the two products $g_{\rm Z'} z_u$ and $g_{\rm Z'} z_q$.
Taking the projected discovery limits in the $\mu^+\mu^-$ channel from
Fig.~\ref{fg:mumu}, they can by translated into limits for the couplings
$g_{\rm Z'} z_u$ and $g_{\rm Z'} z_q$. Fig.~\ref{fg:zuzq} shows
the discovery reach of a collider operating at $\sqrt{s} = 1000$ GeV and ${\cal
L} = 1000 {\rm \ fb}^{-1}$ in terms of the couplings $g_{\rm Z'} z_u$
and $g_{\rm Z'} z_q$ for different $Z'$ masses. Even for $Z'$ masses
that are substantially smaller than the center-of-mass energy, an impressive
sensitivity for small $Z'$ couplings can be achieved (for comparison, the
Standard Model SU(2)$_{\rm L}$ coupling to left-handed quarks is
$|g I_3^q| \approx 0.32$).

The discovery of a $Z'$ boson could be translated into a band in the [$g_{\rm
Z'} z_u$, $g_{\rm Z'} z_q$]-plane
with the shape of the contours in Fig.~\ref{fg:zuzq}. The two quantum numbers
couplings $g_{\rm Z'} z_u$ and $g_{\rm Z'} z_q$ could be
disentangled by examining in addition the branching ratios or the angular
distribution of the decay products of the $Z'$ boson.

\section{Heavy $Z'$ bosons and comparison with hadron colliders}

For $Z'$ masses beyond the center-of-mass energy $\sqrt{s}$, the discovery
sensitivity in the process $e^+e^- \to f\bar{f}$ drops quickly. In this case,
the presence of the $Z'$ boson modifies the signal cross-section only through
off-shell propagator effects. Nevertheless, for sufficiently strong
$Z'f\bar{f}$ couplings, a $Z'$ boson can be discovered indirectly for masses
much larger than $\sqrt{s}$. In Ref.~\cite{rie:00} it has been shown for various
grand unified models that the sensitivity of a linear collider 
extends to $Z'$ masses which are about an
order of magnitude larger than the center-of-mass energy.

For heavy gauge bosons, with $\MZp - \sqrt{s} \gg \GZp$, the contribution to
the amplitudes for the process $e^+e^- \to f\bar{f}$ essentially does not
depend on the s-channel momentum transfer, i.e. the $f\bar{f}$ invariant mass.
Therefore the invariant mass spectrum of the final-state fermions is not a good
observable for searching for heavy gauge bosons. Instead one can look for
deviations in the integrated cross-section. Since in general the contribution
of the $Z'$ boson depends on the initial- and final-state helicities, it is
useful to consider not only the total cross-section, but also the
forward-backward, left-right and polarization asymmetries \cite{rie:00}.
Nevertheless, for simplicity, in the following only the total cross-section for
$e^+e^- \to f\bar{f}$ will be considered.

\begin{figure}[p]
\epsfig{figure=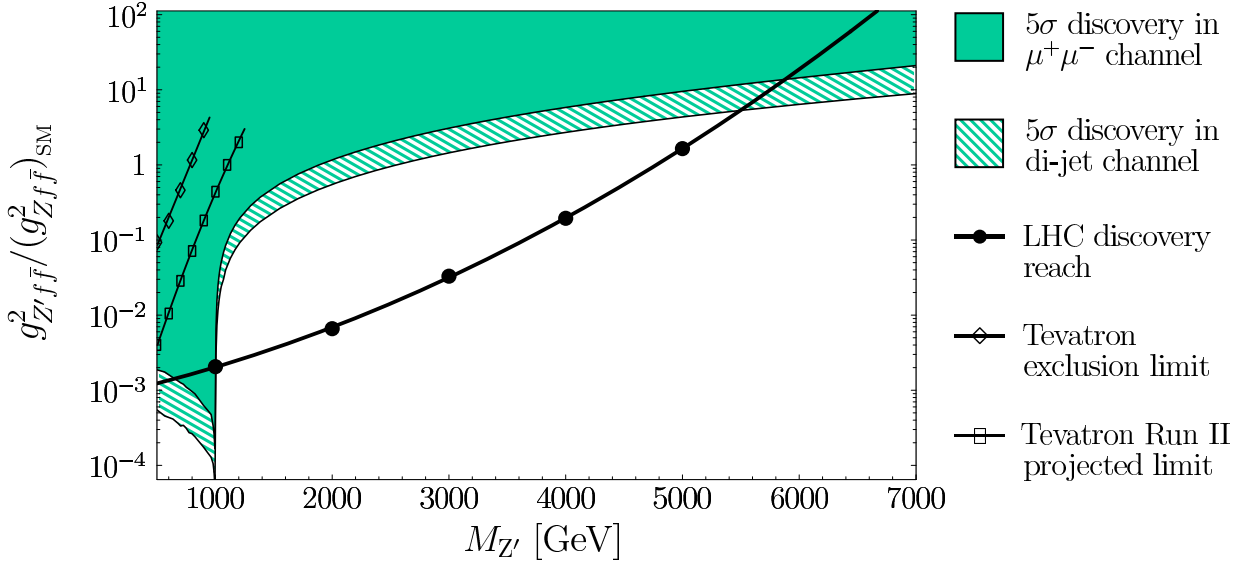, width=\textwidth}
\vspace{-3ex}
\caption{Projected discovery reach of a future $e^+e^-$ collider with $\sqrt{s}
= 1000$ GeV and ${\cal L} = 1000 {\rm \ fb}^{-1}$ for a sequential $Z'$ boson in the
$\mu^+\mu^-$ and the di-jet channels. The depicted range of $Z'$ masses
includes direct and off-resonance searches (see text).
Also shown are the discovery reach of the
LHC \cite{lhc} in the best channel $Z' \to e^+e^-$, the present
limit from searches at the Tevatron \cite{TeVcur} and
the expected exclusion reach at the end of Tevatron Run II.}
\label{fg:Zsm}
\end{figure}
\begin{figure}[p]
\epsfig{figure=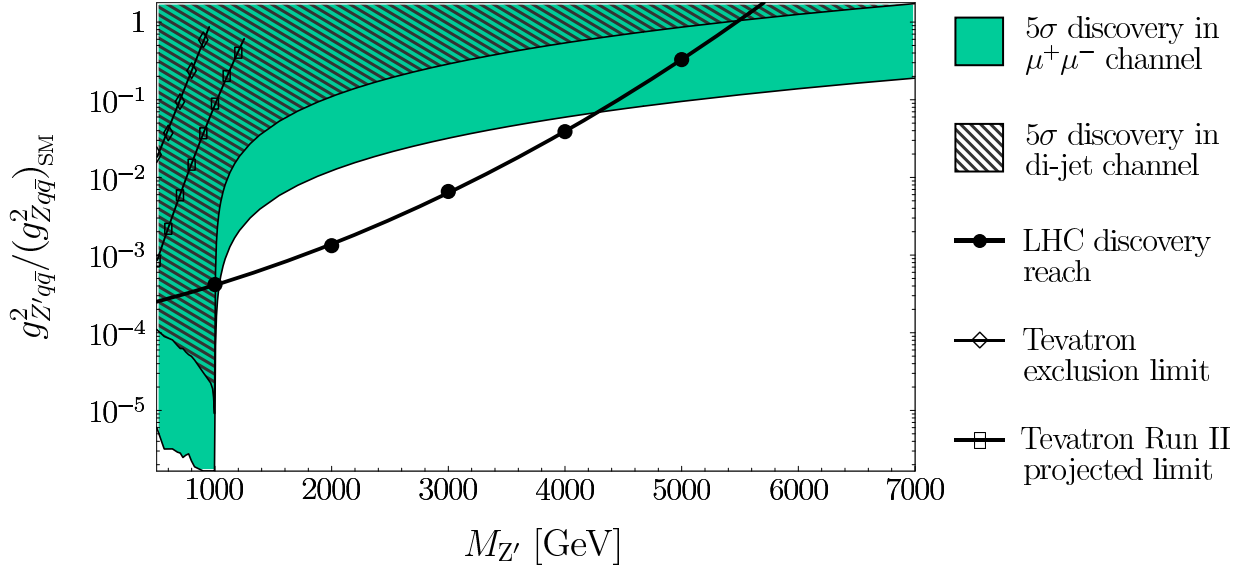, width=\textwidth}
\vspace{-3ex}
\caption{Projected discovery reach of a future $e^+e^-$ collider with $\sqrt{s}
= 1000$ GeV and ${\cal L} = 1000 {\rm \ fb}^{-1}$ for a $Z_{\rm B-L}$ boson in the
$\mu^+\mu^-$ and the di-jet channels. 
Also shown are the discovery reach of the
LHC \cite{lhc} in the best channel $Z_{\rm B-L} \to e^+e^-$, the present
limit from searches at the Tevatron \cite{TeVcur} and
the expected exclusion reach at the end of Tevatron Run II.}
\label{fg:ZBL}
\end{figure}
These off-resonance searches for heavy $Z'$ bosons can be combined with the
direct searches discussed above. Fig.~\ref{fg:Zsm} depicts the $5\sigma$
discovery limits for a $Z'$ boson with couplings that are proportional to the
Standard Model $Z$ boson ("sequential $Z'\,$") in terms of the gauge coupling
$g_{Z'f\bar{f}}$ to the fermions, normalized to the Standard Model coupling.
For $\MZp < \sqrt{s}$ the limits are derived from direct searches as in
Figs.~\ref{fg:mumu} and \ref{fg:qq}, while indirect searches using off-shell
effects in the process $e^+e^- \to f\bar{f}$ provide constraints for $\MZp >
\sqrt{s}$. 
Also shown are the coverage of
the LHC \cite{lhc} and the Tevatron \cite{TeVcur} 
(for better visibility, the data
points have been interpolated by smooth curves). 
The strongest constraints at hadron colliders are obtained from the decay $Z' \to
e^+e^-$, since this channel offers the best momentum resolution \cite{lhc2}.
The current limit from $Z'$ searches at the Tevatron excludes $Z'$ bosons with a
production cross-section times branching ratio of more than $\sim 40$ fb
\cite{TeVcur}, leading to the exclusion contour in Fig.~\ref{fg:Zsm}. 
This limit is expected to improve to $\sim 2$ fb during the
Run II of Tevatron, assuming an integrated luminosity of 4 fb$^{-1}$.

As can be seen from the figure, the linear collider provides very good
sensitivity for $Z'$ bosons with masses below $\sqrt{s} = 1$ TeV and small
couplings, and for very large values of $\MZp$, but relatively strong
couplings. In the intermediate region 1 TeV $< \MZp \lesim$ 5.5 TeV, the coverage
of the LHC is superior, since here $Z'$ bosons of this mass range can still be
directly produced.

From a theoretical point-of-view, a more interesting case than the sequential
$Z'$ is a gauge boson with quantum numbers proportional to the difference
between baryon number $B$ and lepton number $L$. It is a special case of the
minimal U(1) models summarized in Tab.~\ref{tab:charges} with $z_q = z_u$. The
salient feature of such a $Z_{\rm B-L}$ boson is the fact that is does not mix
with the Standard Model $Z$ boson, so that constraints from $Z$-pole precision
data on the $Z_{\rm B-L}$ boson are particularly weak. As pointed out before,
in the absence of mixing in the tree-level mass matrix, mixing effects
generated by higher-order loop contributions can always be rotated away for
on-shell gauge bosons. The discovery reaches of a linear collider and the LHC
for a $Z_{\rm B-L}$ boson are shown in Fig.~\ref{fg:ZBL}, in terms of the $Z'$
gauge coupling to the quarks, $g_{Z'q\bar{q}}$, normalized to the Standard
Model $Z$ coupling. Since a $Z_{\rm B-L}$ boson has a much larger branching
ratio into leptons than a sequential $Z'$, the $\mu^+\mu^-$ channel instead of
the di-jet channel provides the best sensitivity for searches at an $e^+e^-$
collider. In the high-mass region the linear collider  can achieve better
limits than the LHC already for $Z'$ masses $\MZp \gesim 4.3$ TeV in this
model.


\section{Parameter determination}

While the discovery potential for a $Z'$ boson of a few TeV is superior at the
LHC compared to a TeV linear collider, the linear collider can yield important
information about the couplings of the $Z'$ boson to the Standard Model
fermions \cite{rie:00} and thus help to reveal the origin of the new gauge
boson. Using the measurement of the $Z'$ mass at the LHC as an input, the
absolute values of the $Z'$-fermion couplings can be extracted through off-shell
propagator effects even for $\MZp \gg \sqrt{s}$. The reader is referred to
Ref.~\cite{rie:00} for more information.

In case of the discovery of a $Z'$ boson with $\MZp < \sqrt{s}$ through the
radiative-return method, the couplings of the new particle can be studied in
detail by analyzing the branching ratios and angular and polarization
asymmetries. The precision for these measurements can be substantially enhanced
by tuning the collider energy to the $Z'$ resonance.
Due to the high luminosity planned for a future linear collider,
this would allow to measure the properties of the $Z'$ boson 
with a level of precision comparable to the $Z$-pole measurements performed at
LEP, even if the cross-section is several orders of magnitude smaller. In
addition, the analyses can be improved by using polarized $e^\pm$ beams. 

A quantitative discussion of the parameter determination is only possible in
the context of a specific model. In the following, the example of a $Z_{\rm
B-L}$ boson with mass $\MZp = 400$ GeV and lepton coupling $\tilde{g}_l =
0.006$ is considered. This model corresponds to a particular case of the
minimal U(1) models listed in Tab.~\ref{tab:charges} with $z_q = z_u$ and
$\tilde{g}_l = 3 g_{\rm Z'} z_q$. As a special feature, the $Z_{\rm B-L}$ boson
has only vector-like couplings to the Standard Model fermions, i.e. the
couplings to left- and right-handed fermions are identical. The subsequent
discussion will focus on the decay channel of the $Z_{\rm B-L}$ boson into muon
pairs.

Using eq.~\eqref{eq:Gam} one obtains for the total width
$\GZp \simeq 0.6$ MeV. The cross-section in the $\mu^+\mu^-$ channel is computed
from eq.~\eqref{eq:onres} to be
\begin{equation}
 (\sigma \times BR) / (\sigma \times BR)_{\rm SM} \simeq 1.2 \times 10^{-3}.
\end{equation}
As evident from Fig.~\ref{fg:mumu}, such a $Z_{\rm B-L}$ boson could be
discovered at an $e^+e^-$ collider running at $\sqrt{s} = 500$ GeV and ${\cal
L} = 500$ fb$^{-1}$ in the $\mu^+\mu^-$ channel, while a $5\sigma$ discovery at
the LHC is not expected to be possible. Even in this case, in addition to the
discovery of the new particle, it is also feasible to obtain valuable
information about its parameters.

Due to the beam energy spread induced by beamstrahlung and initial-state
radiation, the relatively small width of the new gauge boson, $\GZp \simeq 0.6$ MeV,
cannot be directly resolved in a scan around the $Z'$ resonance.  
As a consequence, absolute branching ratios and coupling parameters can only be
determined by measuring all decay channels of the $Z'$ boson, using photon
tagging for the invisible decay modes as pointed out in section~\ref{inv}.
While this method offers interesting perspectives for recognizing
even exotic decay channels, the precision is limited due to the requirement of a
transverse photon.
It will be further explored at the end of this section.

Nevertheless, it is possible to determine ratios of couplings with high precision.
For example, the ratio of the left-handed and right-handed coupling of the $Z'$
boson to the electrons can be obtained by measuring the production cross-section
for different polarizations of the incoming electron beam,
\begin{equation}
\left( \frac{g^{\rm L}_{Z'ee}}{g^{\rm R}_{Z'ee}} \right)^2 = 
\frac{(P+1)\sigma_{\rm L} + (P-1)\sigma_{\rm R}}{(P-1)\sigma_{\rm L} +
(P+1)\sigma_{\rm R}},
\end{equation}
where $\sigma_{\rm L/R}$ is the cross-section for left-/right-polarized
electrons and $P$ is the polarization degree. In the special case of the
$Z_{\rm B-L}$ boson the polarized cross-sections are identical, $\sigma_{\rm L}
= \sigma_{\rm R}$, so that the determination of $g^{\rm L}_{Z'ee}/g^{\rm
R}_{Z'ee}$ is affected by the uncertainty in the polarization degree $P$
only through statistical fluctuations. The resulting error
is estimated to be of the order of $10^{-4}$ and therefore negligible.
The background contributions from Standard Model sources can either be computed
using theoretical calculations or they can be obtained from measurements with a
center-of-mass energy sufficiently far away from the $Z'$ resonance, which
are then extrapolated to $\sqrt{s} = \MZp$. In both cases the systematic
error of the background subtraction is relatively small and will not
be considered henceforth.

In the given example, using 10 fb$^{-1}$ each for left- and right-polarized
$e^-$ at $\sqrt{s} = 400$ GeV and assuming a polarization degree of $P = 80$\%
would result in $N_{\rm sig} \sim 22500$ signal events over a background of
$N_{\rm bkgd} \sim 4700$ events for left-handed polarization and
$N_{\rm bkgd} \sim 4000$ for right-handed polarization. The
statistical error in the determination of the chiral electron couplings of the
$Z_{\rm B-L}$ boson is then estimated to be
\begin{equation}
\frac{\delta(g^{\rm L}_{Z'ee}/g^{\rm
R}_{Z'ee})}{g^{\rm L}_{Z'ee}/g^{\rm
R}_{Z'ee}} \simeq 0.009. \label{eq:gLR}
\end{equation}
The ratio of left- and right-handed couplings can also be extracted from the
forward-backward asymmetry $A_{\rm FB}$. It is defined as
\begin{equation}
A_{\rm FB} = \frac{\sigma_{\rm F}-\sigma_{\rm B}}{\sigma_{\rm F}+\sigma_{\rm B}},
\qquad \sigma_{\rm F} = \int_0^{\cos\theta_{\rm max}} {\rm d}\cos\theta \,
  \frac{\rm d\sigma}{\rm d\cos\theta},
\quad \sigma_{\rm B} = \int_{-\cos\theta_{\rm max}}^0 {\rm d}\cos\theta \,
  \frac{\rm d\sigma}{\rm d\cos\theta},
\end{equation}
where $\theta$ is the scattering angle between the incoming $e^-$ and the
outgoing fermion $f$. Neglecting the mass of the final-state fermions,
the forward-backward asymmetry for $e^+e^- \to Z' \to f\bar{f}$ 
is expressed through the $Z'$ couplings as follows,
\begin{equation}
A_{\rm FB} = \frac{[(g^{\rm L}_{Z'ee})^2 - (g^{\rm R}_{Z'ee})^2]
	[(g^{\rm L}_{Z'ff})^2 - (g^{\rm R}_{Z'ff})^2]}%
  {[(g^{\rm L}_{Z'ee})^2 + (g^{\rm R}_{Z'ee})^2]
  	[(g^{\rm L}_{Z'ff})^2 + (g^{\rm R}_{Z'ff})^2]} \times
  \frac{\cos\theta_{\rm max}}{1+\frac{1}{3}\cos\theta_{\rm max}}.
  \label{eq:afb}
\end{equation}
Considering the muon channel, i.e. $f=\mu$, and assuming lepton universality,
the four couplings in eq.~\eqref{eq:afb} are reduced to two independent
couplings. Thus the ratio $g^{\rm L}_{Z'll}/g^{\rm R}_{Z'll} = g^{\rm L}_{Z'ee}/g^{\rm R}_{Z'ee} = g^{\rm
L}_{Z'\mu\mu}/g^{\rm R}_{Z'\mu\mu}$ can be determined from the forward-backward
asymmetry in $e^+e^- \to \mu^+\mu^-$ without using polarization
up to a twofold ambiguity,
\begin{equation}
\rule{0mm}{0mm}\hspace{-0.7em}
\left( \frac{g^{\rm L}_{Z'll}}{g^{\rm R}_{Z'll}} \right)^2 = 
  \frac{\sqrt{C} - \sqrt{A_{\rm FB}}}{\sqrt{C} + \sqrt{A_{\rm FB}}}
  \quad \mbox{or} \quad
\left( \frac{g^{\rm L}_{Z'll}}{g^{\rm R}_{Z'll}} \right)^2 = 
  \frac{\sqrt{C} + \sqrt{A_{\rm FB}}}{\sqrt{C} - \sqrt{A_{\rm FB}}},
  \quad \mbox{with} \quad
C = \frac{\cos\theta_{\rm max}}{1+\frac{1}{3}\cos\theta_{\rm max}}.
\end{equation}
For the example of a $Z_{\rm B-L}$ boson, using 20 fb$^{-1}$ and
$\cos\theta_{\rm max} = 0.94$ would result in
$N_{\rm sig} \sim 22500$ signal events each over a background of
$N_{\rm bkgd} \sim 6500$ events for forward scattering and
$N_{\rm bkgd} \sim 2200$ for backward scattering. This leads to the following
statistical error for the determination of the $Z_{\rm B-L}$ couplings,
\begin{equation}
\frac{\delta(g^{\rm L}_{Z'll}/g^{\rm
R}_{Z'll})}{g^{\rm L}_{Z'll}/g^{\rm
R}_{Z'll}} \simeq 0.14. \label{eq:gFB}
\end{equation}
This relatively poor precision is in part a result of the pathological case of
the $Z_{\rm B-L}$ boson, which does not create any forward-backward asymmetry.
The precision can be greatly improved by using the information about the
$Z_{\rm B-L}$-electron couplings from the polarized cross-section measurement
\eqref{eq:gLR} as an additional input. In this case it is also not necessary to
assume lepton universality. With this method a rather impressive 
statistical accuracy for the
determination of the $Z_{\rm B-L}$-muon couplings is achieved,
\begin{equation}
\frac{\delta(g^{\rm L}_{Z'\mu\mu}/g^{\rm
R}_{Z'\mu\mu})}{g^{\rm L}_{Z'\mu\mu}/g^{\rm
R}_{Z'\mu\mu}} \simeq 0.01. \label{eq:gFB2}
\end{equation}
The propagation of the error in $g^{\rm L}_{Z'ee}/g^{\rm R}_{Z'ee}$ from
\eqref{eq:gLR} leads to a negligible error in $g^{\rm L}_{Z'\mu\mu}/g^{\rm
R}_{Z'\mu\mu}$ of the order of $10^{-4}$, since for the $Z_{\rm B-L}$ coupling
structure, this effect enters only through statistical fluctuations.

It is worth mentioning that, in addition to the asymmetry measurements, the
ratio of the left- and right-handed couplings of the $Z'$ boson to tau leptons
can also be extracted from an analysis of the tau polarization, which can be
extracted from the angular distribution of the hadronic decay products.

In a similar way, the ratios of the couplings to different fermion types, e.g.
the ratio of the couplings to leptons and quarks, $g_{Z'll}/g_{Z'qq}$, can be
determined at the per-cent level by comparing the cross-sections for lepton and
jet pair production. If the $Z'$ boson is not a $Z_{\rm B-L}$ boson, the
coupling measurement \eqref{eq:gLR} also depends on the uncertainty in the
polarization degree. This uncertainty could be eliminated  by using the Blondel
scheme \cite{blondel}, if the positron beam can also be polarized.

Next the measurement of the mass of a narrow $Z'$ boson will be
discussed. Due to the limited energy resolution of the trackers and
calorimeters, the most precise determination is achieved by measuring the
cross-section at various center-of-mass energies around the $Z'$ mass.
While it is not possible to directly resolve the resonance line-shape of the
$Z'$ boson, as pointed out above, it is still possible to determine the mass of
the boson accurately if the shape of the beam energy spectrum is precisely
known. The effects of initial-state radiation have been calculated theoretically
to high orders in perturbation theory \cite{LLold,LLnew}, so that the remaining
theoretical uncertainty is negligible. The beamstrahlung effects due to
interactions between the two incoming beams can be either obtained from
numerical simulations based on theoretical models \cite{guinea-pig} or directly
measured using Bhabha scattering \cite{moenig:00}.

\begin{figure}[tb]
\raisebox{7cm}{(a)}\hspace{-3ex}
\epsfig{figure=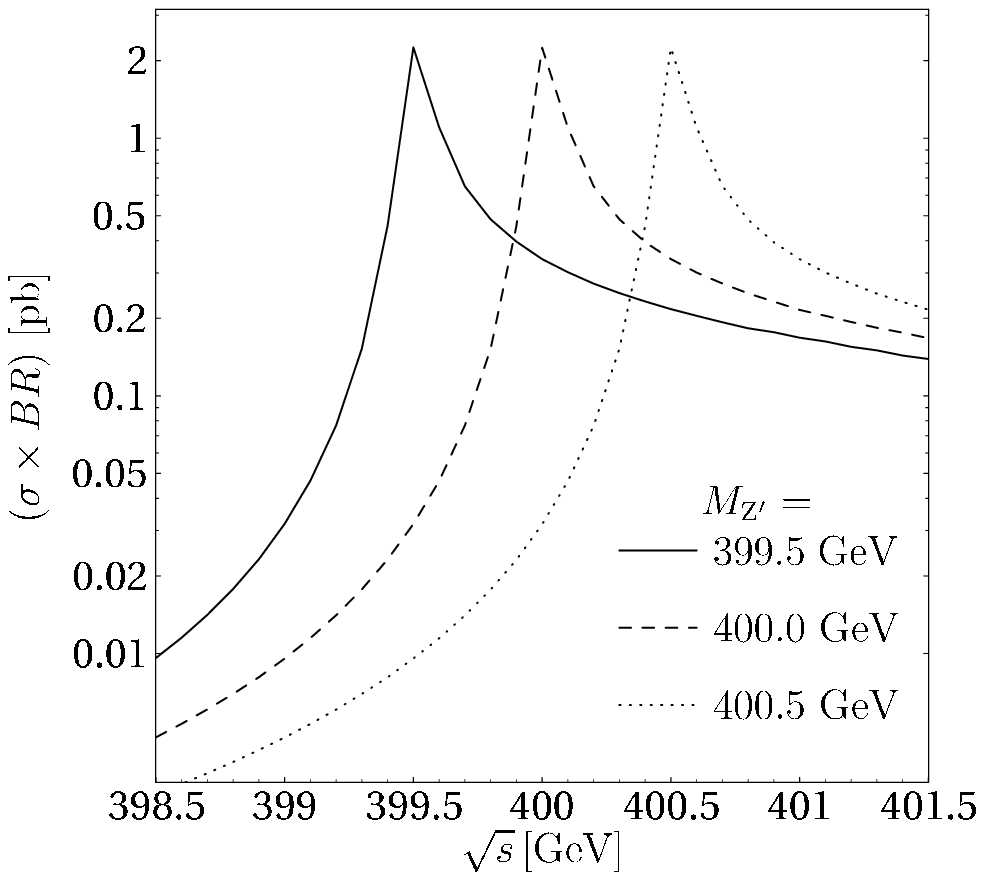, width=8cm, bb=20 432 290 682}
\hspace{1ex}
\raisebox{7cm}{(b)}\hspace{-3ex}
\epsfig{figure=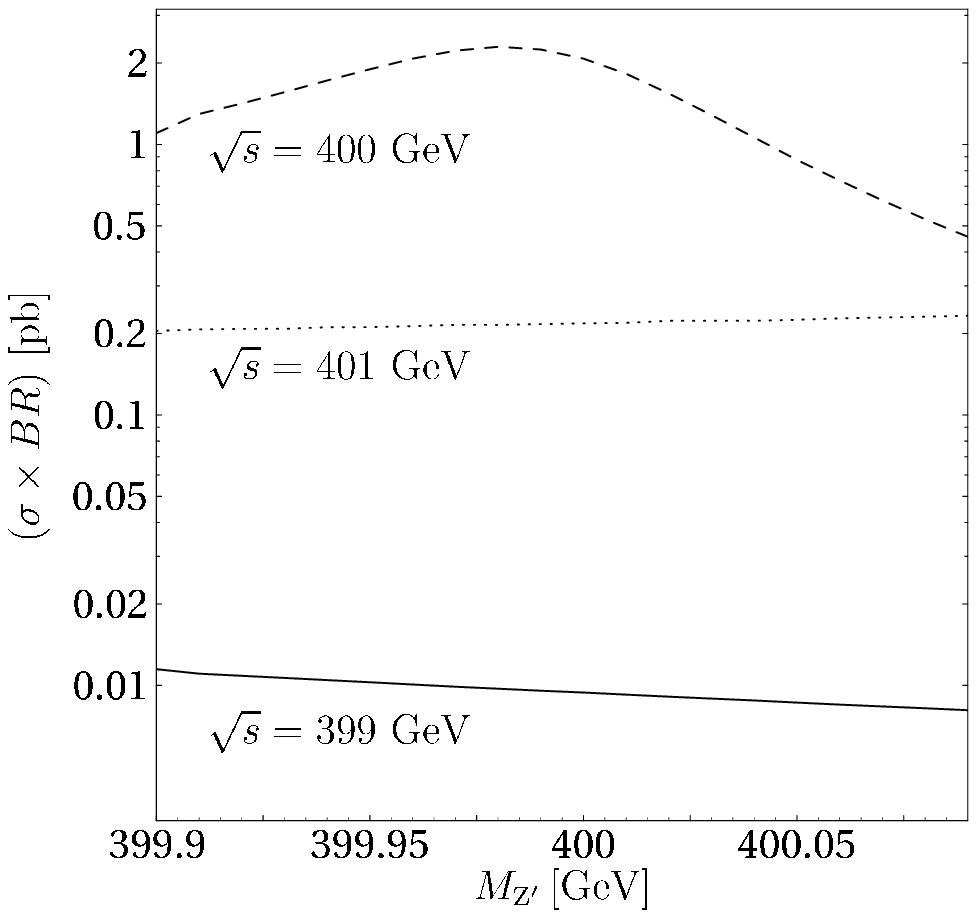, width=8cm, bb=20 421 300 681}
\caption{(a) Dependence of the cross-section for $e^+e^- \to Z_{\rm B-L} \to
\mu^+\mu^-$ on the center-of-mass energy near the resonance peak for a $Z_{\rm
B-L}$ boson with lepton coupling $\tilde{g}_l = 0.006$ and different values of
the mass. (b) Dependence of the cross-section for $e^+e^- \to Z_{\rm B-L} \to
\mu^+\mu^-$ on the mass of the $Z_{\rm B-L}$ boson for three different
center-of-mass energies.}
\label{fg:massscan}
\end{figure}

Including beamstrahlung and initial-state radiation, Fig.~\ref{fg:massscan}~(a)
shows the cross-section near the resonance peak for the $Z_{\rm B-L}$
model and three different values of the mass. As evident from the figure, the
cross-section drops quickly for $\sqrt{s} < \MZp$, but is smeared out for
$\sqrt{s} > \MZp$ due to beamstrahlung and initial-state radiation.
Thus by combining cross-section measurements below and above the
nominal resonance peak, a strong sensitivity on the $Z'$ mass is achieved.

As an example, the mass measurement of the $Z_{\rm B-L}$ boson with $\MZp =
400$ GeV from three measurements of the cross-section $e^+e^- \to (Z_{\rm B-L})
\to \mu^+\mu^-$ at $\sqrt{s} = 399, 400, 401$ GeV is studied. The cross-section
is computed including beamstrahlung and initial-state radiation and the cuts
outlined in section~\ref{direct}. The dependence of the cross-sections on the
mass at the three center-of-mass energies is depicted in
Fig.~\ref{fg:massscan}~(b). Due to the beam energy spread, the cross-section at
$\sqrt{s} = 401$ GeV, i.e. above the mass peak, depends only mildly on the $Z'$
mass. Nevertheless, it is useful to include this point as an absolute
normalization for the other two other scan points, which depend more strongly on
$\MZp$.

\begin{figure}[tb]
\centering
\epsfig{figure=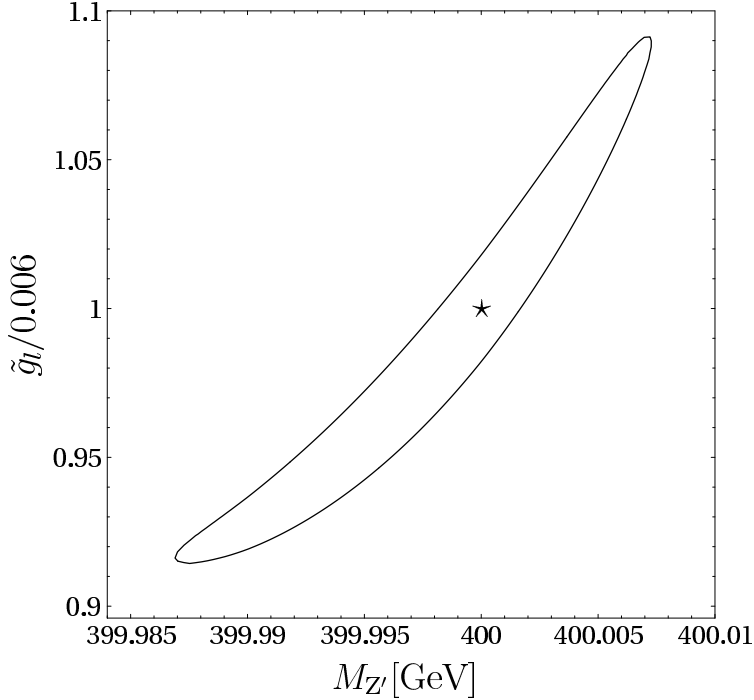, width=10cm, bb=19 445 274 682}
\caption{One sigma confidence level contours for the determination of the mass,
$\MZp$, and the lepton coupling, $\tilde{g}_l$, of a $Z_{\rm B-L}$ boson with $\MZp =
400$ GeV and $\tilde{g}_l = 0.006$, obtained from simulated measurements of the
cross-section $e^+e^- \to (Z_{\rm B-L}) \to \mu^+\mu^-$ at three center-of-mass
energies, $\sqrt{s} = \MZp$, $\MZp-1$ GeV, $\MZp+1$ GeV. The star indicates the
values of the underlying model. The vertical axis for the leptonic coupling,
$\tilde{g}_l$, is normalized to its nominal value of $\tilde{g}_l = 0.006$.
In practice, due to the unknown value of the total $Z_{\rm B-L}$ width, this
coupling could be determined only in arbitrary units.}
\label{fg:coup}
\end{figure}

It is assumed that 10 fb$^{-1}$ is spent at each of the
three center-of-mass energies. The mass is derived from the cross-section
measurements by using a binned $\chi^2$ fit with the mass and the lepton
coupling, $\tilde{g}_l$, as unknown parameters (the unknown value of the total
width can be absorbed into $\tilde{g}_l$). In Fig.~\ref{fg:coup} the resulting
one sigma contours are shown.
The beam energy spread leads to a strong correlation between the mass and the
coupling of the $Z'$ boson. 
From the end-points of the contour in Fig.~\ref{fg:coup}, the following
estimate for the statistical errors of the mass determination is obtained,
\begin{equation}
\MZp = 400.0^{+0.007}_{-0.013} {\rm \ GeV.}
\end{equation}
At this high level of precision, the mass measurement error is dominated by
systematic uncertainties. The absolute value of the beam energy is expected to
be controllable at the level of $10^{-4}$ or better \cite{tesladet}, leading to
an error of 40 MeV in the $Z'$ mass. An additional uncertainty arises from the
beamstrahlung spectrum.
The beamstrahlung spectrum can be expressed through a simple parametrization
introduced in Ref.~\cite{circe},
\begin{equation}
G_{\rm beam}(x) = a_0 \, \delta(1-x) + a_1 \, x^{a_2} (1-x)^{a_3},
\label{eq:beampar}
\end{equation}
where $x$ is the fraction of the beam energy in the collision process. From
measurements of Bhabha events, the parameters $a_i$ can be precisely determined,
yielding for {\sc Tesla} design parameters \cite{moenig:00}
\begin{equation}
\begin{array}{@{}r@{\;}c@{\;}l@{}}
a_0 &=& 0.5274 \pm 0.0014 \\
a_2 &=& 13.895 \pm 0.082 \\
a_3 &=& -0.6314 \pm 0.0021 \\
\end{array}
\qquad \mbox{correlation matrix: }
\left(
\begin{array}{lll}
1.000 & 0.430 & 0.708 \\
0.430 & 1.000 & 0.755 \\
0.708 & 0.755 & 1.000 \\
\end{array}
\right). \label{eq:beamerr}
\end{equation}
The fourth parameter $a_1$ is given by the normalization condition $\int_0^1
{\rm d}x \, G_{\rm beam}(x) =1$. The influence of the uncertainty of the
beamstrahlung spectrum on the mass determination has been estimated by folding
the parametrization \eqref{eq:beampar} with the $Z'$ resonance function, which
has been approximated by a $\delta$-function. Taking the errors and correlations
in \eqref{eq:beamerr}, a rather small effect on the extracted mass of 1 MeV
is obtained. Other error sources like the uncertainty in the determination of
the luminosity of the selection efficiency have been checked to be negligible.
The total error is dominated by the error in the beam energy determination,
resulting in
\begin{equation}
\MZp = 400.0^{+0.041}_{-0.042} {\rm \ GeV.}
\end{equation}
Thus one can conclude from this simplified study that a mass measurement with a
precision at the level of $10^{-4}$ seems feasible.

If the couplings of a hypothetical $Z'$ boson are stronger, it might already be
discovered at the Tevatron Run II or the LHC. Nevertheless, in this case
measurements at the linear collider can supply important information about the
couplings and mass of the extra gauge boson. Besides increasing the precision
with respect to measurements at hadron colliders, the linear collider also can 
investigate observables that are inaccessible at hadron colliders.

In particular, if the total decay width of the $Z'$ boson is of the same order
of magnitude as the average beam energy loss due to initial-state radiation and
beamstrahlung (which is a few GeV for $\sqrt{s} \sim 500$ GeV), it may be
directly resolved in a scan around the $Z'$ resonance. As a consequence, the
measurement not only of \emph{relative}, but
of \emph{absolute} branching ratios would be possible.

Alternatively, absolute branching ratios can also be determined by making use
of photon tagging for the invisible decay channels, as pointed out above. The
precision of this method is limited due to the requirement of a sufficiently
hard transverse photon. As a consequence there is no enhancement due to the
large collinear logarithms $L = \log s/\me^2$ as for the visible decay
channels. In addition, because the tagging photon requires some minimum energy,
the collider energy cannot be tuned directly to the $Z'$ resonance. Therefore
it is not possible to determine total branching ratios of very weakly coupled
$Z'$ bosons, such as the example of a $Z_{\rm B-L}$ with lepton coupling
$\tilde{g}_l = 0.006$.

For moderate coupling strength, however, the measurement of the invisible decay
channel is feasible. In the following a $Z_{\rm B-L}$ boson with the same mass
as before, $\MZp = 400$~GeV, but larger lepton coupling $\tilde{g}_l = 0.1$
will be considered. The signal is characterized by a photon plus missing
energy, $e^+e^- \to \gamma + \Eslash$. The photon is required to have a minimum
energy of $E_\gamma^{\rm min} = 10$~GeV and to be emitted at an angle with
respect to the beam axis larger than $\theta_\gamma^{\rm min} = 20^\circ$. 

For a $Z_{\rm B-L}$ boson, the invisible decay modes originate from decays into
neutrinos.
The invariant mass $M_{\nu\bar{\nu}}$ of the neutrino pair can be deduced from
the photon energy $E_\gamma$, $M_{\nu\bar{\nu}}^2 = s ( 1- E_\gamma/\sqrt{s})$.
Accordingly, the resolution of the invisible mass spectrum is governed by the 
photon energy resolution of the electromagnetic calorimeter, which is expected
to be \cite{tesladet}
\begin{equation}
\Delta E_\gamma/E_\gamma = 10\%/\sqrt{E_\gamma/{\rm GeV}} \oplus 1\%.
\label{eq:egbin}
\end{equation}
As before, the photon energy spectrum
is divided into bins of the size given by eq.~\eqref{eq:egbin}, with the signal
concentrated in a single bin.

It turns out that the signal $e^+e^- \to Z_{\rm B-L} \gamma \to
\nu\bar{\nu}\gamma$ is maximized for a collider energy of about $\sqrt{s}
\approx 420$~GeV, resulting in $N_{\rm sig} \sim 5700$ events for 100 fb$^{-1}$
luminosity. The main background arises from the Standard Model contributions to
$e^+e^- \to \nu\bar{\nu}\gamma$, which are mediated by s-channel $Z$-boson
exchange or t-channel $W$-boson exchange. For the bin size eq.~\eqref{eq:egbin},
about $N_{\rm bkgd} \sim 20000$ background events are expected. This leads to a
statistical uncertainty in the determination of the ``invisible'' $Z'$
cross-section of 2.8\%.

The main systematic uncertainties for this measurement arise from the
parametric uncertainty of the $W$-boson mass in the Standard Model background
and the modeling of possible conversion of the photon into $e^+e^-$ pairs in
the detector, leading to a total systematic error of 0.4\% \cite{invZ}. In
addition, there is a theoretical uncertainty arising from the subtraction of the
background. It is assumed that the Standard Model background can be calculated
with an accuracy of 0.2\%, which induces an error of 1.1\% on the signal
cross-section. Adding statistical and systematic uncertainties, the total error
for the measurement of the ``invisible'' $Z'$ cross-section in this
example is 3.1\%.

Combining the measurement of the invisible cross-section with measurements of
the visible decay modes of the $Z'$ boson, i.e. the leptonic and hadronic
cross-sections, allows to determine absolute branching ratios.  In principle,
one could also use the same photon tagging method for the visible decay
channels as for the invisible decays, thereby minimizing the systematic
uncertainties and theoretical assumptions. However, a more precise
determination of the visible decay modes can be achieved by tuning the collider
energy to the $Z'$ resonance, as described before. In fact the error of
resonance measurements is negligible compared to the error of the  invisible
cross-section determination from $e^+e^- \to Z_{\rm B-L} \gamma \to
\nu\bar{\nu}\gamma$. The combination of the visible cross-section measurements
on the resonance and the invisible cross-section measurement via photon tagging
requires the inclusion of  beamstrahlung and photon radiation spectra. These
spectra can be controlled with a precision far better than 1\%, leading to a
negligible uncertainty for the branching ratios.

The resulting errors for the absolute branching
ratios are summarized in Tab.~\ref{tab:BR}. For the decay modes into leptons
and light quarks, flavor universality has been assumed. As evident from the
table, although the neutrino branching ratio in the given example can only be
determined with an error of a few percent, the precision for the other
branching ratios is better than 1\%.

\begin{table}[tb]
\renewcommand{\arraystretch}{1.2}
\centering
\begin{tabular}{|c|r@{$\,\pm\,$}l|c|}
\hline
Decay channel & \multicolumn{2}{c|}{Simulated measurement} & Relative error \\
\hline
$l^+l^-$ & $\qquad$0.469 & 0.0035 & 0.8\% \\
$q\bar{q}$, $q \neq t$ & 0.261 & 0.002 & 0.8\% \\
$t\bar{t}$ & 0.035 & 0.0003 & 0.8\% \\
$\nu\bar{\nu}$ & 0.235 & 0.0055 & 2.5\% \\
\hline
\end{tabular}
\caption{Expected precision for the determination of absolute branching ratios for a
$Z_{\rm B-L}$ boson with $\MZp = 400$~GeV and lepton coupling $\tilde{g}_l =
0.1$. The leptonic branching ratio has been derived under assumption of lepton
universality.}
\label{tab:BR}
\end{table}


\section{Conclusions}

In summary, the prospects of a future high-luminosity linear collider for
searches for a new neutral $Z'$ gauge boson have been reexamined, for the case
that the $Z'$ boson decays (primarily) into charged Standard Model fermions
$f$. As a novel feature, the capability of a linear collider for direct
detection of narrow $Z'$ resonances through radiative return to the $Z'$ pole
has been investigated. These direct searches provide an unparalleled
sensitivity for weakly coupled $Z'$ bosons, allowing the discovery for
$Z'f\bar{f}$ that are up to two orders of magnitude weaker than the couplings
of the Standard Model $Z$ boson. In comparison with the LHC, a linear collider
provides the best coverage for $Z'$ searches for masses $\MZp$ below the
$e^+e^-$ center-of-mass energy $\sqrt{s} \lesim 1$ TeV, while the reach of the LHC is
superior for higher masses in the range 1 TeV $< \MZp \lesim$ 5 TeV. For very
large masses $\MZp \gg 1$ TeV, the linear collider can again achieve
competitive limits through indirect effects of off-shell $Z'$ bosons in the
process $e^+e^- \to f\bar{f}$.

This study provides a simple estimate of the capabilities of a future linear
collider for $Z'$ searches, without including experimental acceptances,
systematic uncertainties or other detector-related effects and taking into
account only the dominant background sources. While these effects could modify
the predictions for the event rates at the order of 10--30\%, the general
picture will nevertheless remain the same in a more refined analysis. 

If a weakly coupled $Z'$ boson with $\MZp \lesim 1$ TeV is discovered either at
hadron colliders or the linear collider, the $Z'$-fermion couplings could be
measured precisely by tuning the collider energy to the $Z'$ resonance. Even if
these couplings are several orders of magnitude smaller than the couplings of
the Standard Model $Z$ boson, couplings ratios and branching ratios could be
determined at the per-cent level. The mass of the $Z'$ boson could be
determined with a relative error of about $10^{-4}$.

Finally, an $e^+e^-$ collider provides unique opportunities for searching for
exotic $Z'$ bosons, for example ``hadrophobic'' $Z'$ bosons, which couple to the
Standard Model leptons, but not to the quarks, or ``invisible'' $Z'$ bosons,
which primarily decay into undetectable particles, e.~g. neutrinos. In the
latter case, the searches are based on
the process $e^+e^- \to \gamma + {\rm missing\ energy}$, using a hard photon 
for tagging.


\section*{Acknowledgments}

\noindent
This work greatly benefited from inspiring ideas and careful proofreading of
the manuscript by M.~Carena, B.~Dobrescu and M.~Schmitt. The author is grateful
to M.~Schmitt and M.~K.~Unel for help on the Tevatron limits
and to S.~Mrenna for useful discussions. This work is supported by
the U.S. Department of Energy under Contract DE-AC02-76CH03000.



\begin{thebibliography}{99}
\frenchspacing

\bibitem{gut}
For a review, see
P.~Langacker,
Phys.\ Rept.\  {\bf 72}, 185 (1981).

\bibitem{tc}
For a recent review, see
C.~T.~Hill and E.~H.~Simmons,
Phys.\ Rept.\  {\bf 381}, 235 (2003).

\bibitem{leike:99}
A.~Leike,
Phys.\ Rept.\  {\bf 317}, 143 (1999).


\bibitem{TeVcur}
D\O\ coll., V.~M.~Abazov {\it et al.}, D\O note 4375-Conf (2004);\\ 
D.~Waters, talk
given at the {\it 19th Annual Lake Louise Winter Institute}, Lake Louise,
Alberta (2004);\\
see also
A.~Dominguez (on behalf of the CDF coll.) and R.~Kehoe (on behalf of the
D\O\ coll.), talks given at the Fermilab Joint Experimental-Theoretical Physics
Seminar, Aug 8, 2003.

\bibitem{bogdan:03}
T.~Appelquist, B.~A.~Dobrescu and A.~R.~Hopper,
Phys.\ Rev.\ D {\bf 68}, 035012 (2003).

\bibitem{lc}
\textsc{Tesla} Technical Design Report, Part III, eds. R.~Heuer,
D.~J.~Miller, F.~Richard and P.~M.~Zerwas, DESY-2001-11C
[hep-ph/0106315];\\
T.~Abe {\it et al.}  [American Linear Collider Working Group Collaboration],
in {\it Proc. of the APS/DPF/DPB Summer Study on the Future of Particle Physics
(Snowmass 2001)}, eds. R.~Davidson and C.~Quigg,
SLAC-R-570
[hep-ex/0106056];\\
K.~Abe {\it et al.}  [ACFA Linear Collider Working Group Collaboration],
KEK-REPORT-2001-11
[hep-ph/0109166].

\bibitem{structf}
E.$\,$A.~Kuraev and V.$\,$S.~Fadin,
Sov.$\,$J.$\,$Nucl.$\,$Phys.$\,${\bf 41} (1985) 466
[Yad.$\,$Fiz.$\,${\bf 41} (1985) 733];\\
G.~Altarelli and G.~Martinelli,
in {\it Physics at LEP}, eds. J.~Ellis and R.~Peccei (CERN-86-02), Vol.~1,
p.~47.

\bibitem{LLold}
F.~A.~Berends {\it et al.}, in
  {\it Z Physics at LEP 1}, eds. G.~Altarelli, R.~Kleiss and
  C.~Verzegnassi (CERN-89-08), p.~89.

\bibitem{LLnew}
M.~Skrzypek and S.~Jadach,
Z.\ Phys.\ C {\bf 49}, 577 (1991);\\
M.~Skrzypek,
Acta Phys.\ Polon.\ B {\bf 23}, 135 (1992);\\
M.~Cacciari, A.~Deandrea, G.~Montagna and O.~Nicrosini,
Europhys.\ Lett.\  {\bf 17}, 123 (1992);\\
M.~Przybycie\'n,
Acta Phys.\ Polon.\ B {\bf 24}, 1105 (1993);\\
G.~Montagna, O.~Nicrosini and F.~Piccinini,
Phys.\ Lett.\ B {\bf 406}, 243 (1997).

\bibitem{circe}
T.~Ohl,
Comput.\ Phys.\ Commun.\  {\bf 101}, 269 (1997).

\bibitem{tesla}
{\sc Tesla} Technical Design Report, Part I, eds. F.~Richard,
J.~R.~Schneider, D.~Trines and A.~Wagner, DESY-2001-11A,
[hep-ph/0106314].

\bibitem{guinea-pig}
D.~Schulte,
CERN-PS-99-014.

\bibitem{nlc}
N.~Phinney ({\it ed.}) {\it et al.} [NLC Collaboration],
in {\it Proc. of the APS/DPF/DPB Summer Study on the Future of Particle Physics
(Snowmass 2001) } eds. R.~Davidson and C.~Quigg,
SLAC-R-571;\\
N.~Akasaka {\it et al.} [JLC Design Study Group],
KEK-REPORT-97-1.

\bibitem{clic}
G.~Guignard ({\it ed.}),
{\it A 3-TeV $e^+e^-$ linear collider based on CLIC technology},
CERN-2000-008.

\bibitem{napoly:92}
O.~Napoly,
CERN-SL-92-28-AP.

\bibitem{wilson:01} G.~Wilson,
in {\it 2nd ECFA/DESY Linear Collider Study} (2001), p. 1498,
LC-PHSM-2001-009
[\texttt{http://www-flc.desy.de/lcnotes/}].

\bibitem{tt}
D.~Peralta, M.~Martinez and R.~Miquel, in {\it Proc. of the International
Workshop on Linear Colliders (LCWS 99)}, Sitges, Spain (1999).

\bibitem{tesladet}
{\sc Tesla} Technical Design Report, Part IV, eds. T.~Behnke,
S.~Bertolucci, R.D.~Heuer and R.~Settles, DESY-2001-011D.

\bibitem{invZ}
M.~Carena, A.~de Gouv\^ea, A.~Freitas and M.~Schmitt,
Phys.\ Rev.\ D {\bf 68}, 113007 (2003).

\bibitem{rie:00} S.~Riemann,
in {\it 2nd ECFA/DESY Linear Collider Study} (2001), p. 1451,
LC-TH-2001-007
[\texttt{http://www-flc.desy.de/lcnotes/}].

\bibitem{lhc}
ALTAS Technical Design Report, CERN/LHCC-99-15, p. 940.

\bibitem{lhc2}
P.~Camarri {\it et al.}, in {\it Proc. of the Large Hadron Collider Workshop},
eds. G.~Jarlskog and D.~Rein, Aachen, Germany (1990), Vol. II, p. 709.

\bibitem{blondel}
A.~Blondel,
Phys.\ Lett.\ B {\bf 202}, 145 (1988)
[Erratum-ibid.\  {\bf 208}, 531 (1988)].

\bibitem{moenig:00} K.~M\"onig
in {\it 2nd ECFA/DESY Linear Collider Study} (2001), p. 1353,
LC-PHSM-2000-060
[\texttt{http://www-flc.desy.de/lcnotes/}].

\end{thebibliography}
\end{document}